# A New Two-Dimensional Functional Material with Desirable Bandgap and Ultrahigh Carrier Mobility

Ning Lu[1, *, ⊥], Zhiwen Zhuo[2, ⊥], Hongyan Guo[1,*], Ping Wu[3], Wei Fa[4], Xiaojun Wu[2], and Xiao Cheng Zeng[5,6, *]

[1] Anhui Province Key Laboratory of Optoelectric Materials Science and Technology, Department of Physics, Anhui Normal University, Wuhu, Anhui, 241000, China.
[2] CAS Key Laboratory of Materials for Energy Conversion, School of Chemistry and Materials Sciences and CAS Center for Excellence in Nanoscience, and Hefei National Laboratory of Physical Sciences at the Microscale, University of Science and Technology of China, Hefei, Anhui 230026, China
[3] School of Physics, The University of Sydney, Sydney, New South Wales 2006, Australia
[4] National Laboratory of Solid State Microstructures and Department of Physics, Nanjing University, Nanjing 210093, China
[5] Department of Chemistry, University of Nebraska-Lincoln, Lincoln, NE 68588, USA
[6] Collaborative Innovation Center of Chemistry for Energy Materials, University of Science and Technology of China, Hefei, Anhui 230026, China

**ABSTRACT** Two-dimensional (2D) semiconductors with direct and modest bandgap and ultrahigh carrier mobility are highly desired functional materials for nanoelectronic applications. Herein, we predict that monolayer $CaP_3$ is a new 2D functional material that possesses not only a direct bandgap of 1.15 eV (based on HSE06 computation), and also a very high electron mobility up to 19930 $cm^2 V^{-1} s^{-1}$, comparable to that of monolayer phosphorene. More remarkably, contrary to the bilayer phosphorene which possesses dramatically reduced carrier mobility compared to its monolayer counterpart, $CaP_3$ bilayer possesses even higher electron mobility (22380 $cm^2 V^{-1} s^{-1}$) than its monolayer counterpart. The bandgap of 2D $CaP_3$ can be tuned over a wide range from 1.15 to 0.37 eV (HSE06 values) through controlling the number of stacked $CaP_3$ layers. Besides novel electronic properties, 2D $CaP_3$ also exhibits optical absorption over the entire visible-light range. The combined novel electronic, charge mobility, and optical properties render 2D $CaP_3$ an exciting functional material for future nanoelectronic and optoelectronic applications.

## Introduction

Two-dimensional (2D) atomic-layered materials, such as graphene, boron-nitride, transition metal dichalcogenides and phosphorene, have attracted enormous interests due to their remarkable properties not seen in their bulk counterparts.[1-10] For example, monolayer phosphorene has been shown to possess novel electronic properties,[1, 11-19] including a desired bandgap of 1.51 eV, as well as high carrier (hole) mobilities in the range of 10000-26000 cm$^2$ V$^{-1}$ s$^{-1}$.[14] Indeed, few-layer black phosphorus field-effect transistors have been successfully fabricated,[13, 18-20] which give rise to drain current modulation up to $10^5$ and hole mobility up to 5200 cm$^2$ V$^{-1}$ s$^{-1}$.[13, 20] Today, the quest for new 2D atomic-layer materials that possess both desired direct bandgap and high carrier mobilities is still a highly active area of research. Recent advances in developing 2D phosphorene-related materials includes new allotropes of phosphorene and 2D phosphides (phosphorene derivatives) with superior properties. In particular, a series of phosphorene derivatives are predicted to possess high carrier mobilities,[21-25] e.g., 2D hittorfene with hole mobility in the range of 3000-7000 cm$^2$ V$^{-1}$ s$^{-1}$, 2D GeP$_3$ with hole and electron mobility of 8480 and 8840 cm$^2$ V$^{-1}$ s$^{-1}$, respectively, as well as 2D InP$_3$ with electron mobility of 1919 cm$^2$ V$^{-1}$ s$^{-1}$. These 2D phosphorene derivatives can be viewed as exfoliated atomic layers from the corresponding bulk materials.

In this communication, we report a new member in the phosphorene derivative family, namely, 2D CaP$_3$ structure with distinctly different structure from recently reported 2D GeP$_3$ and 2D InP$_3$.[21,22] This new atomic-layered material possesses several fascinating electronic properties compared to previously reported phosphorene derivatives. Note that the bulk CaP$_3$, discovered in 1973,[26] is a metal phosphide. It belongs to a Ca-P compound consisting of porous framework of phosphorene with Ca atoms located at P dimer vacancy site. Recently, the MP$_3$ (M = Ca, Sr, Ba) series are predicted to be topological nodal-line semimetals.[27] Here, our comprehensive density-functional theory (DFT) computations suggest that 2D monolayer CaP$_3$ is a direct-gap semiconductor with a bandgap of 1.15 eV (based on HSE06 functional). The bandgap of 2D multi-layered CaP$_3$ ranges from 1.15 to 0.37 eV (HSE06), depending on the number of layers. Also importantly, 2D CaP$_3$ exhibits strongly anisotropic carrier mobilities with the electron mobility being as high as 22380 cm$^2$ V$^{-1}$ s$^{-1}$. The latter value is comparable the predicted hole mobility of phosphorene.

## Results and discussion

**Structure and Stability**

The optimized structures of bulk and monolayer CaP$_3$ are shown in Figure 1. Bulk CaP$_3$ exhibits *P-1* symmetry and its crystalline structure can be viewed as layered porous framework of phosphorene with Ca atoms located at P dimer vacancy site. CaP$_3$ monolayer is constructed by taking an atomic layer from the CaP$_3$ bulk in the (0 0 1) direction. Monolayer CaP$_3$ also exhibits *P-1* symmetry as the bulk. However, some reconstructions of the structure are worthy of noting, e.g., the distortion of the P-P bond, and the positions of the Ca atoms being closer to the plane of the CaP$_3$ sheet, rendering the thickness of the atomic layer thinner for about 0.49 Å. The lattice constants of the CaP$_3$ monolayer are $a$ = 5.59 Å, $b$=5.71 Å, and $\gamma$ = 81.09 °, slightly different from those of the bulk lattice. CaP$_3$ monolayer also exhibits a puckered configuration, similar to black phosphorene. In CaP$_3$ monolayer, the P-Ca bond length is 2.83 - 2.97Å, while the P-P bond length is 2.21-2.24 Å, shorter than that in the phosphorene (2.24 - 2.28 Å). Electron localization function shows clear ionic bonding between Ca-P (see Figure S1). Bader charge analysis suggests that each Ca atom transfers about 1.39 *electron* to P atoms (see Table S2). For the CaP$_3$ bilayer and trilayer, our DFT calculation shows that both prefer the same stacking order as bulk, and the optimized structure of the bilayer CaP$_3$ is shown in Figure S2.

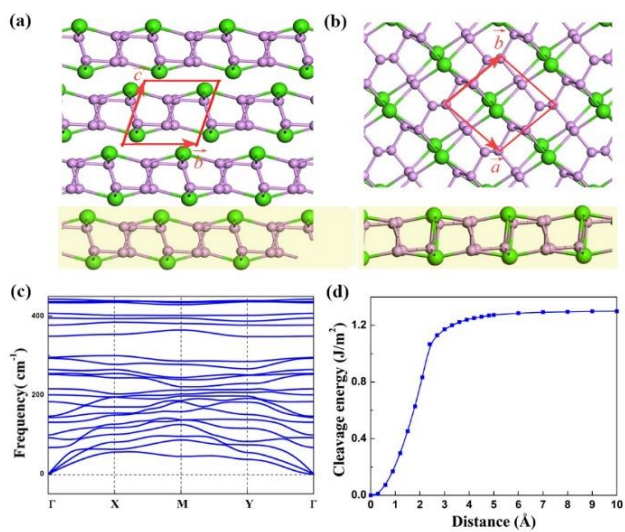

**Figure 1.** (a) Top and side views of optimized bulk CaP$_3$, and (b) CaP$_3$ monolayer structures. The unit cell is marked in red line. The purple and green spheres represent P and Ca atoms, respectively.

(c) Phonon dispersion curves of the CaP$_3$ monolayer. (d) Cleavage energy $E_{cl}$ in J/m$^2$ versus the separation distance $d$ for a planar fracture within the bulk CaP$_3$.

The dynamic stability of CaP$_3$ monolayer is confirmed via computing the phonon dispersion curves, which show no imaginary phonon modes (Figure 1c). The highest vibration frequency is 443 cm$^{-1}$, comparable to that of phosphorene (450 cm$^{-1}$)[28] and 2D GeP$_3$ (480 cm$^{-1}$),[21] reflecting the mechanical robustness of the covalent P-P bonds. Thermal stability of structures is also examined via BOMD simulations in which the temperature of the system is controlled at 300, 600, and 1000 K, respectively. The overall structure is still well kept after 5 ps simulation even when the temperature is up to 1000 K (see Figure S4). We note that the use of periodic boundary condition in BOMD simulation with relatively small system size can artificially increase the stability of the structure. Still, the simulation at the elevated temperature (1000 K) can suggest that the CaP$_3$ monolayer is highly stable, at least at the ambient temperature.

To examine easiness in mechanical exfoliation of a CaP$_3$ monolayer from the bulk, we compute the cleavage energy $E_{cl}$, characterized by the interlayer coupling strength.[29] To this end, a planar fracture is introduced within a unit cell of CaP$_3$ bulk. The CaP$_3$ supercell with four layers is used so that the interaction between two neighboring fractures due to periodic boundary condition can be neglected. As shown in Figure 1(d), the calculated cleavage energy $E_{cl}$ is 1.30 J/m$^2$. We have also computed the cleavage energy taken for the separation of a CaP$_3$ monolayer from a trilayer. Similar result is obtained. By comparing the cleavage energy of 1.30 J/m$^2$ for bulk CaP$_3$ with several other atomic-layered materials, e.g., bulk Ca$_2$N (1.08 J/m$^2$),[30] GeP$_3$ (1.14 J/m$^2$)[21] and InP$_3$ (1.32 J/m$^2$),[22] we conclude that it should be quite feasible to exfoliate a monolayer CaP$_3$ from the bulk.

**Electronic Structure**

The computed electronic structure of monolayer CaP$_3$ is shown in Figure 2a, where the CaP$_3$ monolayer exhibits a direct bandgap of 0.69 eV (at the PBE level). The more accurate calculations with the HSE06 functional and the hybrid PBE0 functional yield a bandgap value of 1.15 eV and 2.0 eV, respectively. From the computed partial density of states (PDOS), one can see that the valence band maximum (VBM) is mainly contributed by the $p$ orbitals of P component, while the conduction band minimum (CBM) is a hybridization of *p* orbitals of P and *d* orbitals of Ca. The partial charge density shown in Figure S5 is consistent with the PDOS analysis.

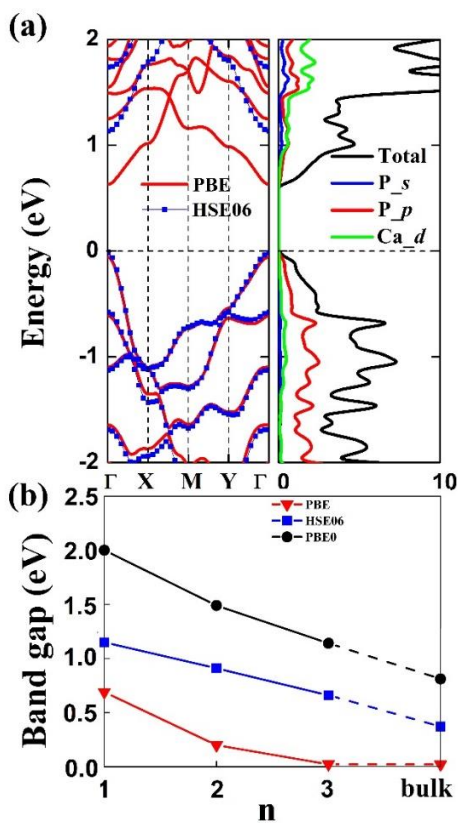

**Figure 2.** (a) Computed band structures, the total and partial density of states (DOS) of $CaP_3$ monolayer, and (b) computed bandgap of $CaP_3$ multilayers versus the number of atomic layers, by using three DFT functionals, i.e., PBE, HSE06, and hybrid PBE0.

For 2D $CaP_3$ multilayers, their electronic structure can be strongly dependent on the number of layers. Also, the HSE06 computation indicates that the 2D $CaP_3$ multilayers with different thickness still retain the direct bandgap feature (Figure S6). As shown in Figure 2(b) and Table S3, the bandgap decreases with increasing the number of atomic layers. The $CaP_3$ bilayer possesses a direct bandgap of 0.91 eV (HSE06), while the trilayer possess a direct bandgap of 0.67 eV (HSE06). The lower limit of the bandgap is that of the bulk whose value is 0.37 eV. These results suggests that the bandgap of $CaP_3$ thin film can be adjusted over a quite large range of 1.15 eV - 0.37 eV (HSE06) through controlling the number of stacked atomic layers. A possible reason for the bandgap decrease with the number of atomic layer is due to the lack of strong interlayer interaction and structure reconstruction upon stacking atomic layers. Detailed band structures and PDOS of CaP$_3$ bilayer, trilayer and bulk are shown in Figure S6.

Note also from Figure 2(b) that all three DFT functional, PBE, HSE06 and hybrid PBE0, give rise to the same general trend of the bandgap reduction with the increase of the number of layers. Hence, we expect that the predicted physical trend in the bandgap change is realistic (a similar conclusion has been drawn for the widely studied phosphorene), and will be confirmed by future experiments.

**Optical Absorption**

Optical properties of 2D CaP$_3$ have also been computed based on the HSE06 functional. The frequency-dependent dielectric function $\varepsilon(\omega) = \varepsilon_1(\omega) + i\varepsilon_2(\omega)$ is calculated, with which the absorption coefficient can be evaluated according to the expression:[31, 32] $\alpha(\omega) = \frac{\sqrt{2}\omega}{c} \{[\varepsilon_1^2(\omega) + \varepsilon_2^2(\omega)]^{\frac{1}{2}} - \varepsilon_1(\omega)\}^{\frac{1}{2}}$. The absorption coefficients, with the electric field vector ***E*** being polarized in parallel to the plane ($\alpha_{//}$) or being perpendicular to the plane ($\alpha_\perp$), are computed. Clearly, the CaP$_3$ monolayer exhibits optical absorption over wide wave-length range in the visible light region (Figure 3). Due to the larger cross section, in-plane absorption is always greater than out-of-plane absorption. The CaP$_3$ bilayer exhibits stronger optical absorption than the monolayer. The outstanding optical properties suggest potential applications of 2D CaP$_3$ as efficient optical absorber materials in solar cells or infrared detection optoelectronic devices.

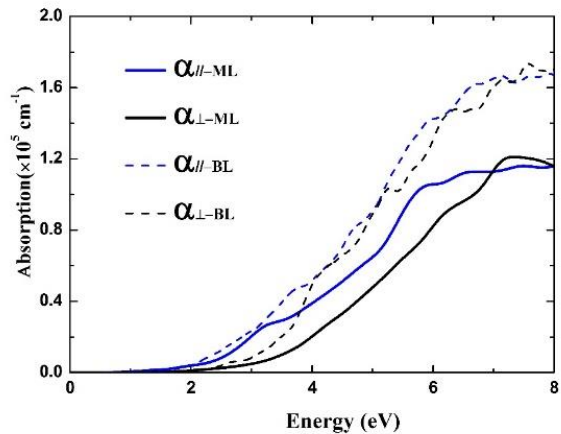

**Figure 3.** Calculated absorption coefficient of CaP$_3$ monolayer and bilayer at the HSE06 level.

**Carrier Mobility**

The carrier mobility of the 2D CaP$_3$ is calculated on the basis of the deformation potential (DP) theory,[33] which has been widely used to predict the carrier mobility of 2D atomic layered structures.[21-23, 34, 35] Note that the DP theory based mobility tends to overestimate the measured mobilities. For phosphorene, the calculated hole mobility based on the DP theory is around 10000 cm$^2$ V$^{-1}$ s$^{-1}$,[14] which overestimates the measured hole mobility of 5200 cm$^2$ V$^{-1}$ s$^{-1}$.[20] In any case, according to the DP theory, the carrier mobility of 2D systems can be described by the following formula:[23] $\mu_{2d} = \frac{e\hbar^3 C_{2d}}{k_B T m^* m_d (E_1)^2}$, where $k_B$ is Boltzmann constant, $T$ is the temperature (300 K), $m^*$ is the effective mass in the transport direction, and $m_d = (m_a^* m_b^*)^{1/2}$ is the average effective mass. Because the directions *a* and *b* for the unit cell of CaP$_3$ monolayer are close to be perpendicular with one another, the $m_d$ is considered to be $(m_a^* m_b^*)^{1/2}$ for simplicity. For the CaP$_3$ bilayer, a convention unit cell is used in the calculation, and the directions *a* and *b* are defined in Figure S2. $C_{2d}$ is the elastic moduli. $E_1 = \frac{\Delta E_{edge}}{\Delta l/l}$ is the deformation potential constant,[23] $\Delta E_{edge}$ is the energy change of the band edge, $l$ is the equilibrium lattice constant, and $\Delta l$ is the change of the lattice constant.

Our results are summarized in Table 1. For the CaP$_3$ monolayer, the elastic moduli are 54.64 J/m$^2$ and 64.38 J/m$^2$ along the *a* and *b* direction, respectively, both being slightly higher than those of P related monolayers (e.g., hittorfene[23] ~ 50 J/m$^2$ and 2D InP$_3$[22] ~ 43 J/m$^2$). The effective mass of electron is nearly the same along the *a* [0.80 $m_e$ (the mass of free electron)] and *b* [0.83 $m_e$] directions. However, the effective mass of hole is anisotropic. The effective mass along the *a* direction (1.07 $m_e$) is notably higher than that along the *b* direction (0.36 $m_e$). Similar to hittorfene and black phosphorene, the deformation potential $E_1$ is anisotropic, with the electron along the *a* direction has the smallest value of about 0.3 eV. This leads to a large electron mobility along the *a* direction, up to 1.99×10$^4$ cm$^2$ V$^{-1}$ s$^{-1}$. It is noted that the electron mobility along the *b* direction also exhibits a large value of ~1.75×10$^3$ cm$^2$ V$^{-1}$ s$^{-1}$. The hole mobility is only ~0.08×10$^3$ and 0.78×10$^3$ cm$^2$ V$^{-1}$ s$^{-1}$ along the *a* and *b* direction, respectively, much lower than that of electron mobility.

**Table 1**. Carrier effective mass (*m**), elastic modules (*C*), deformation potential constant (*E$_1$*), and carrier mobility (*μ*) of the CaP$_3$ monolayer (ML) and bilayer (BL).

|  | $m^*/m_e$ | $C$ (J m$^{-2}$) | $E_1$ (eV) | $\mu$ ($10^3$ cm$^2$V$^{-1}$s$^{-1}$) |
| --- | --- | --- | --- | --- |
| ML-Electron (*a*) | 0.80 | 54.6 | 0.30 | 19.9 |
| ML-Electron (*b*) | 0.83 | 64.4 | 1.08 | 1.75 |
| ML-Hole (*a*) | 1.07 | 54.6 | 4.59 | 0.08 |
| ML-Hole (*b*) | 0.36 | 64.4 | 2.81 | 0.78 |
| BL-Electron (*a*) | 0.83 | 87.2 | 0.35 | 22.4 |
| BL-Electron (*b*) | 0.81 | 120.7 | 2.78 | 0.50 |
| BL-Hole (*a*) | 0.81 | 87.2 | 3.39 | 0.57 |
| BL-Hole (*b*) | 0.15 | 120.7 | 10.0 | 0.49 |

**Discussion**

Among all phosphorene derivatives reported so far, 2D CaP$_3$ possesses the highest electron mobility while phosphorene possesses the highest hole mobility. 2D CaP$_3$ can be excellent complementary material for phosphorene in electron mobility. And electrons in 2D CaP$_3$ are much more mobile than hole, contrary to the trend for phosphorene and hittorfene, for which hole exhibits higher mobility. For phosphorene, the carrier mobility decreases rapidly from its monolayer to bilayer. In contrast, the CaP$_3$ bilayer even possesses a higher electron mobility, up to 2.24×10$^4$ cm$^2$ V$^{-1}$ s$^{-1}$, along the *a* direction, indicating the superior optical absorption with high mobility can be achieved by a thicker CaP$_3$ multilayer.

Compared to recently reported 2D GeP$_3$ or InP$_3$,[21, 22] despite of apparent similarity in chemical formula, the structure of CaP$_3$ is fundamentally different. Specifically, the structure of 2D GeP$_3$ and InP$_3$ can be regarded as a semi-single-atom-thick sheet with combination of single metal unit and $P_6$ hexatomic ring. There is no continuous framework of P in the two structures, but only $P_6$ cluster. And the coordination of all atoms is 3. In contrast, in the 2D crystalline structure of $CaP_3$ the coordination of Ca and most P atoms is more than 3. The structure is anisotropic, resulting in anisotropic electronic and mechanical properties. So the dramatically different structure, electronic properties of 2D $CaP_3$ are fundamentally different from those of 2D $GeP_3$ or $InP_3$, as shown previously. In particular, electron mobilities of 2D $GeP_3$ and $InP_3$ are an order of magnitude lower than that of 2D $CaP_3$ ($10^4$ cm$^2$ V$^{-1}$ s$^{-1}$). In addition, $CaP_3$ monolayer possesses a direct bandgap of 1.15 eV (HSE06), higher than that of $GeP_3$ (indirect, 0.55 eV (HSE06)) and $InP_3$ (indirect, 1.14 eV (HSE06)). So 2D $CaP_3$ would be more suitable for electronic and optoelectronic applications than 2D $GeP_3$.

## Conclusion

In conclusion, a new 2D semiconducting $CaP_3$ is predicted to possess novel electronic properties, compared with other state-of-the-art members in the 2D material family. Most notably, $CaP_3$ monolayer exhibits ultra-high electron mobility up to $1.99 \times 10^4$ cm$^2$ V$^{-1}$ s$^{-1}$ while its bilayer possesses even higher electron mobility of $2.24 \times 10^4$ cm$^2$V$^{-1}$s$^{-1}$ than the monolayer. The bandgap of 2D $CaP_3$ decreases with increasing the atomic-layer number, indicating that the bandgap of 2D $CaP_3$ can be adjusted over an energy range from 1.15 eV (monolayer, HSE06) to 0.37 eV (bulk, HSE06), by controlling the number of atomic layers. Both $CaP_3$ monolayer and bilayer exhibit strong optical absorption in the visible-light region. Phonon dispersion curves and BOMD simulation confirm dynamic and thermal stability of $CaP_3$ monolayer. And the cleavage energy of bulk $CaP_3$ suggests high feasibility in exfoliation of a monolayer from the bulk. Overall, 2D $CaP_3$ represents a new class of 2D material with remarkable electronic properties that hopefully can be confirmed by experiments in near future.

## Computational methods

All calculations are performed within the framework of DFT, implemented in the Vienna *ab initio* simulation package (VASP 5.3).[36, 37] The generalized gradient approximation (GGA) in the form of the Perdew-Burke-Ernzerhof (PBE) functional, and projector augmented wave (PAW) potentials are used.[38-40] The valence electron of P is $2s^22p^3$, and Ca is $3s^23p^64s^2$. Dispersion-corrected DFT method (optB88-vdW functional) is used in all the structure optimization,[41, 42] which has been proven reliable for phosphorene systems.[14] The vacuum spacing in the supercell is larger than 15 Å so that interaction among periodic images can be neglected. An energy cutoff of 500 eV is adopted for the plane-wave expansion of the electronic wave function. During the DFT calculation, the *k*-point sampling is carefully examined to assure that the calculation results are converged. Geometry structures are relaxed until the force on each atom is less than 0.01 eV/Å, while the energy convergence criteria of $1 \times 10^{-5}$ eV are met. For each system, the supercell is optimized to obtain the lattice parameters of the lowest-energy structure. For bulk $CaP_3$, the optB88-vdW computation predicts the lattice constants of *a* = 5.60 Å, *b* = 5.68 Å, and *c* = 5.62 Å, in good agreement with experimental values of *a* = 5.59 Å, *b* = 5.67 Å, and *c* = 5.62 Å.[26] Since DFT/GGA method tends to underestimate bandgap of semiconductors, the screen hybrid HSE06 method is also used to examine the band structures.[43] With the optB88-vdW optimized structure, the computed bandgap of the bulk $CaP_3$ at the HSE06 level of theory is 0.37 eV, consistent with the previous study.[27] For phosphorene, the HSE06 functional tends to underestimate the electronic band gap but give optical gaps very close to the experiment values, while the more computationally demanding hybrid PBE0 functional gives an electronic band gap closer to the experiment. In any case, the PBE0 computation is also performed.[44] Detailed discussion of different DFT methods is given in Supplemental Information.

The Bader's atom in molecule (AIM) method (based on charge density topological analysis) is used for the charge population calculation.[45] The BOMD simulations are performed using the CASTEP 7.0 package,[46,47] for which the supercell contains 96 atoms. First, the PBE functional and ultrasoft pseudopotential are selected. The energy cut-off is 280 eV. The BOMD simulations are carried out in the constant-temperature and constant-pressure ensemble. The temperature (300 K, 600 K, or 1000 K) and pressure (1 atm) are controlled using the Nosé-Hoover[48] and Anderson-Hoover method,[49] respectively. The time step in BOMD simulations is 1 fs. Each independent BOMD simulation lasts 5 ps. Moreover, to confirm dynamics stability of the 2D structure, the phonon dispersion spectrum is also computed using the CASTEP 7.0 package, for which the finite displacement method is selected.

**Supporting Information**

Lattice constant, bandgap, band structure and PDOS of $CaP_3$ bulk and 2D layers. Charge population analysis, ELF, partial charge density distribution of VBM and CBM, BOMD snapshots, energy shift of VBM and CBM with respect to the strain of $CaP_3$ monolayer. Structure view and phonon band dispersions of $CaP_3$ bilayer.

**Corresponding Authors**

*N.L., Email: luning@ ahnu.edu.cn;

*H.G., Email: hyguowd@ ahnu.edu.cn;

*X.C.Z., Email: xzeng1@unl.edu

**Author Contributions**

N.L. and Z.Z. contribute equally to this work.

**Notes**

There are no conflicts to declare.

ACKNOWLEDGMENTS

This work was supported by the National Natural Science Foundation of China (No. 21503002, No. 21403005 and No. 11474150); and Anhui Provincial Natural Science Foundation (No.1608085QB40). X.C.Z. was supported by US NSF through the Nebraska Materials Research Science and Engineering Center (MRSEC) (grant No. DMR-1420645), and by a State Key R&D Fund of China (2016YFA0200604) to USTC for summer research, and UNL Holland Computing Center.